\documentstyle[preprint,aps]{revtex}

\begin{document}
\draft
\tighten

%%%%%%%%%%%%%%%%
%%%% Title Page
%%%%%%%%%%%%%%%%

\preprint{\tighten\vbox{\hbox{\bf CLNS 99-1630}
                        \hbox{\bf CLEO 99-10}
			\hbox{\date{\today}}}}

\title{Measurement of the product branching fraction \\
            ${\cal B}(c \to \Theta_c X) \cdot
            {\cal B}(\Theta_c \to \Lambda X$)}

\author{CLEO Collaboration}

\date{\today}

\maketitle
\tighten
%%%%%%%%%%%%%%
%%%% Abstract
%%%%%%%%%%%%%%

\begin{abstract}
Based on an a high statistics $e^+e^-\to c{\overline c}$ 
data sample, % of 4.7 $fb^{-1}$, 
we report on the
inclusive rate for charmed baryons to
decay into $\Lambda$ particles using charm-event
tagging.  We select $e^+e^- \to c\overline{c}$ events
which have a clear anti-charm tag 
and measure the $\Lambda$
content in the hemisphere opposite the tag 
(charge conjugate modes are implicit).
This allows us to determine the product branching
fraction: 
${\cal B}_\Lambda = {\cal B}(c \to \Theta_c X)
\cdot {\cal B}(\Theta_c \to \Lambda X$), where $\Theta_c$
represents a sum over all charmed baryons produced in
$e^+e^-$ fragmentation at $\sqrt{s}$=10.5 GeV, given our specific tags.  
We obtain
${\cal B}_\Lambda$ = (1.87 $\pm$ 0.03 $\pm$ 0.33)\%.
\end{abstract}


\newpage

\begin{center}
R.~Ammar,$^{1}$ P.~Baringer,$^{1}$ A.~Bean,$^{1}$ D.~Besson,$^{1}$,
P.~Brabant,$^{1}$
H.~Bull,$^{1}$ R.~Davis,$^{1}$ J.~Holliday,$^{1}$ S.~Kotov,$^{1}$ 
I.~Kravchenko,$^{1}$
% ...
\end{center}
 
\small
\begin{center}
$^{1}${University of Kansas, Lawrence, Kansas 66045}\\
% ...
\end{center}
\setcounter{footnote}{0} 

%%%%%%%%%%%%%%%%%%
%%%% Introduction
%%%%%%%%%%%%%%%%%%

\section{Introduction}
\label{sec:introduction}

Inclusive measurements of charmed baryon decay products provide 
essential information on the relative contributions of different decay
processes (e.g., external W-emission, internal W-emission, W-exchange)
to the weak $c\to sW^+$ transition in baryons. 
Difficulties in distinguishing direct charm decay products from 
jet fragmentation particles
have hampered such inclusive measurements.
An example is the measurement of ${\cal B}(\Lambda_c \to
\Lambda X)$ (although $``\Lambda_c''$ here designates
$\Lambda_c^+$, charge conjugation is implicit throughout). 
Using the total $\Lambda$ yield at $\sqrt{s}$
= 10 GeV/c to measure ${\cal B}(\Lambda_c \to \Lambda X)$ requires separating
the $\Lambda$ component due to light quark 
fragmentation from production via $c \to
\Lambda_c \to \Lambda X$.  
The difficulty of separating fragmentation
$\Lambda$'s from those resulting from $\Lambda_c$ decays can be
overcome by using a tagged sample of $e^+e^-\to c{\overline c}$ 
events.  In this analysis, we
use charm-event tagging to
measure the product of the
likelihood for a charm quark to 
materialize as a charmed baryon $\Theta_c$ %(${\cal B}(c\to\Theta_c)$)
times the branching fraction for a charmed baryon to decay into
a $\Lambda$: 
${\cal B}(c\to\Theta_c X)\cdot {\cal B}(\Theta_{c} \to \Lambda 
X)$. From JETSET 7.3 
Monte Carlo simulations\onlinecite{lund92}, using the default LEP-tuned
control
parameters, 
we expect that 
$\sim$88\% of
$\Theta_c$ particles produced in $e^+e^-\to c{\overline c}$
fragmentation at $\sqrt{s}$=10 GeV will be $\Lambda_c$'s.

Previous measurements
of ${\cal B}(\Lambda_c\to\Lambda X)$ have either measured the
increase in the $\Lambda$ production rate as the 
$e^+e^-\to\Lambda_c{\overline\Lambda_c}$ threshold is crossed\cite{hybr86}, 
topologically tagged $\Lambda_c$ decays\cite{emul87},
or derived
a value for ${\cal B}(\Lambda_c\to\Lambda X)$ based on measurements of baryon
production in B-meson decay\cite{cleo92}. 
Knowing, for example, that 
${\cal B}(B\to p/{\overline p}(direct)+
anything)=5.5\pm0.5$\%\onlinecite{pdg98},
${\cal B}(B\to \Lambda/{\overline \Lambda}+anything)=4.0\pm0.5$\%,
and assuming that 
${\overline B}\to\Lambda_c{\overline N}$X 
dominates baryon production in B-decay, with
${\overline N}$ equally likely to be ${\overline p}$ or ${\overline n}$, 
%then we have ${\overline B}\to\Lambda_c{\overline p}$X = 
%${\overline B}\to\Lambda_c{\overline n}$X. Then,
one can estimate:
${\cal B}(\Lambda_c\to\Lambda X)\sim{4\over 5.5\times 2}$=36\%. This
simple-minded 
estimate, however, needs to be modified to take into account many corrections,
among them the recent result that
${{\cal B}({\overline B}\to{\overline\Lambda} X)\over {{\cal B}(\overline
B}\to\Lambda X)}=0.43\pm0.09\pm0.07$\cite{Flavor-tag}. 
That latter result implies that
only $\sim$2/3 of the inclusive ($\Lambda+{\overline\Lambda}$) yield 
in ${\overline B}$-decay come from decays of charmed baryons; the remainder
is presumably due to associated production or decay of anticharm baryons.
%${\overline B}\to\Lambda_c(\to\Lambda X){\overline N}$X (assuming that
%charmed baryons produced in B-decay are dominantly $\Lambda_c$).
In this Article, we use a new technique to determine the product of the
probability for a charm quark to produce a charmed baryon $\Theta_c$ times
the probability that the charmed baryon will decay into a $\Lambda$:
${\cal B}_\Lambda = {\cal B}(c \to \Theta_c X)
\cdot {\cal B}(\Theta_c \to \Lambda X$), using continuum $e^+e^-$ 
annihiliation events at $\sqrt{s}$=10.55 GeV.

A sample c\=c event is schematically shown below, showing some of the
particles relevant to our measurement. In the event depicted below,
the
fragmentation of the original c\=c quark-antiquark results in a
$\Lambda$ recoiling in one hemisphere opposite the anti-charm tag 
(either the soft pion or the electron) in
the other hemisphere. For this analysis, event ``hemispheres'' are
defined using the axis which minimizes the momentum (charged plus
neutral) transverse
to that axis (the ``thrust'' axis).
Note that both the soft pion and the electron are of charge opposite to
the $p$ daughter of the $\Lambda$. 
In addition to the tags depicted below,
we also tag c\=c events with fully reconstructed
${\overline D^0}\to K^+\pi^-$ or $D^-\to K^+\pi^-\pi^-$ events, in which
the ${\overline D}$ daughter
kaon contains an \=s-quark, in contrast to the
$\Lambda$. For all four tags,
we will therefore
refer to our signal as an opposite hemisphere, opposite sign (OH/OS)
correlation.

\begin{center}
\[
   \begin{array}{cccccccc}
      & & & c & \overline{c} & & & \\
      & & \Lambda_c \hookleftarrow & & & \hookrightarrow
          D^{*-} & & \\
      & \Lambda X \hookleftarrow & & & & & \hookrightarrow
          \pi^-_{soft} + \overline{D^0} & \\
      p + \pi^- \hookleftarrow & & & & & & & \hookrightarrow
          e^- + \overline{\nu}_e + K^+ \\
   \end{array}
\]
\end{center}

We note that the $\Lambda_c$ in this example can, in practice,
be any charmed baryon and will from here on be denoted as ``$\Theta_c$''.
%As mentioned above,
%replacement of the $D^{*-}$ in the above
%diagram by $D^-\to K^+\pi^-\pi^-$ or replacement of the
%${\overline D^0}\to  e^- + \overline{\nu}_e + K^+$ with
%${\overline D^0}\to\pi^-K^+$ also constitutes a valid tag.

%%%%%%%%%%%%%%%%%%%%%%%%%%%%%%%%%%%
%%%% Apparatus and Event Selection
%%%%%%%%%%%%%%%%%%%%%%%%%%%%%%%%%%%

\section{Apparatus and Event Selection}
\label{sec:event_selection}

This analysis was performed using the CLEO II detector operating at the
Cornell Electron Storage Ring (CESR) at center-of-mass energies $\sqrt{s}$
= 10.52--10.58 GeV.  
The CLEO II detector is a general purpose solenoidal magnet
spectrometer and calorimeter designed to trigger efficiently on two-photon,
tau-pair, and hadronic events \onlinecite{kubota92}.  
Measurements of charged particle momenta are made with
three nested coaxial drift chambers consisting of 6, 10, and 51 layers,
respectively.  These chambers fill the volume from $r$=3 cm to $r$=1 m, with
$r$ the radial coordinate relative to the beam (${\hat z}$) axis. 
This system is very efficient ($\epsilon\ge$98\%) 
for detecting tracks that have transverse momenta ($p_T$)
relative to the
beam axis greater than 200 MeV/c, and that are contained within the good
fiducial volume of the drift chamber ($|\cos\theta|<$0.94, with $\theta$
defined as the polar angle relative to the beam axis). 
This system achieves a momentum resolution of $(\delta p/p)^2 =
(0.0015p)^2 + (0.005)^2$ ($p$ is the momentum, measured in GeV/c). 
Pulse height measurements in the main drift chamber provide specific
ionization resolution
of 5.5\% for Bhabha events, giving good $K/\pi$ separation for tracks with
momenta up to 700 MeV/c and separation of order 2$\sigma$ in the relativistic
rise region above 2 GeV/c. 
Outside the central tracking chambers are plastic
scintillation counters, which are used as a fast element in the trigger system
and also provide particle identification information from time 
of flight
measurements.  

Beyond the time-of-flight system is the electromagnetic calorimeter,
consisting of 7800 thallium-doped CsI crystals.  The central ``barrel'' region
of the calorimeter covers about 75\% of the solid angle and has an energy
resolution which is empirically found to follow:
\begin{equation}
\frac{ \sigma_{\rm E}}{E}(\%) = \frac{0.35}{E^{0.75}} + 1.9 - 0.1E;
                                \label{eq:resolution1}
\end{equation}
$E$ is the shower energy in GeV. This parameterization includes
effects such as noise, and translates to an
energy resolution of about 4\% at 100 MeV and 1.2\% at 5 GeV. Two end-cap
regions of the crystal calorimeter extend solid angle coverage to about 95\%
of $4\pi$, although energy resolution is not as good as that of the
barrel region. 
The tracking system, time of flight counters, and calorimeter
are all contained 
within a superconducting coil operated at 1.5 Tesla. 
Flux return and tracking
chambers used for muon detection are located immediately outside the coil and 
in the two end-cap regions.

The event sample 
used for this measurement is comprised of 3.1 $fb^{-1}$ of data
collected at the $\Upsilon$(4S) resonance and 1.6 $fb^{-1}$ of data 
collected about 60 MeV below the $\Upsilon$(4S) resonance. Approximately
$5\times 10^6$ continuum c\=c events are included in this sample.
For our analysis, we
select continuum hadronic events which 
contain either a low-momentum pion $\pi^-_{soft}$ emitted
at small angles relative to the event thrust axis (from 
$D^{^*-}\to{\overline D^0}\pi^-_{soft}$),
a high momentum electron (from ${\overline c}\to{\overline s}e^-\nu_e$), or
a fully reconstructed ${\overline D}^0\to K^+\pi^-$ or $D^-\to K^+\pi^-\pi^-$
as a tag of $e^+e^-\to c{\overline c}$ events.  
In order to suppress background and enrich the
hadronic fraction of our event sample, we impose several event requirements. 
Candidate events must have: (1) at least five
detected, good quality, charged tracks; (2) an event vertex consistent with
the known $e^+e^-$ interaction point; (3) a total measured visible 
event energy, equal to the sum of the 
observed charged plus neutral energy
$E_{vis} (= E_{chrg} + E_{neutral}$), greater than 110\% of the single
beam energy, $E_{vis}$ $>$ 1.1 $\cdot$ $E_{beam}$.  In addition, when using
an electron to tag a c\=c event we require that either the beam energy
$E_{beam}$ be less than 5.275 GeV or that the event be well collimated.
Specifically, the ratio of Fox-Wolfram event shape parameters $H2/H0$ 
can be used to quantify the ``jettiness'' of an event\cite{FoxWolf}.
For a perfectly
spherical flow of event energy, this ratio equals 0; for a perfectly 
jetty event, this ratio equals 1.0. For our electron tags,
we require this ratio to be greater than 0.35.
This requirement is necessary to remove contamination from 
B${\overline{\rm B}}$ events. Similarly, when using 
${\overline D^0}\to K^+\pi^-$ or
$D^-\to K^+\pi^-\pi^-$ as our charm tags, we eliminate
B${\overline{\rm B}}$ background by requiring that the reconstructed
${\overline D}$ momentum exceed 2.3 GeV/c.

%%%%%%%%%%%%%%%%%%%%%%%%
%%%% Tag Identification
%%%%%%%%%%%%%%%%%%%%%%%%

\section{Tag Identification}
\label{sec:tag_id}

\subsection{Electron Tags}

To suppress background from fake electrons, as well as 
true electrons not necessarily
associated with $e^+e^- \to$ c\=c events, we require that our electron-tag
candidates satisfy the following criteria:

(a) The electron must pass a strict ``probability of electron''
identification criterion. This identification likelihood 
combines measurements of 
a given track's specific ionization deposition in the central drift chamber
with the ratio of the energy of the associated calorimeter shower to the
charged track's momentum\cite{ELECTRON-ID-REF}. 
True electrons have shower energies approximately
equal to their drift chamber momenta;
hadrons tend to be minimum ionizing and have considerably
smaller values of shower energy relative to their measured momenta.
We require that the logarithm of the
ratio of a charged track's
electron probability relative to the probability that the charged track
is a hadron ${\tt L}_e$ be greater than 7 (${\tt L}_e\ge 7$).

(b) The momentum of the electron must be greater than 1 GeV/c.  This 
criterion
helps eliminate kaon and pion fakes and also suppresses electrons from
photon conversions
($\gamma \to e^+ e^-$) and $\pi^0$ Dalitz
decays ($\pi^0 \to \gamma e^+ e^-$).

(c) The electron must have an impact parameter relative to the event vertex
less than 4 mm along the radial coordinate and no more than 2 cm along
the beam axis. This provides additional suppression of electrons resulting
from photon conversions.

%We note that, according to Monte Carlo simulations based on JETSET 7.3,
%electrons which pass these criteria tag a c\=c sample which is $>$ 98\% pure.

\subsection{Soft pion tags}
Our soft pion tag candidates must pass the following restrictions:

(a) The pion must have an impact parameter relative to the event vertex
less than 5 mm along the radial coordinate and no more than 5 cm along
the beam axis.

(b) The pion must pass a 99\% probability criterion for pion identification,
based on the associated 
charged track's specific ionization measured in the drift chamber.

(c) The pion's measured momentum must be between 0.15 GeV/c and 0.40 GeV/c.

\begin{figure}
\begin{picture}(200,250)
\includegraphics{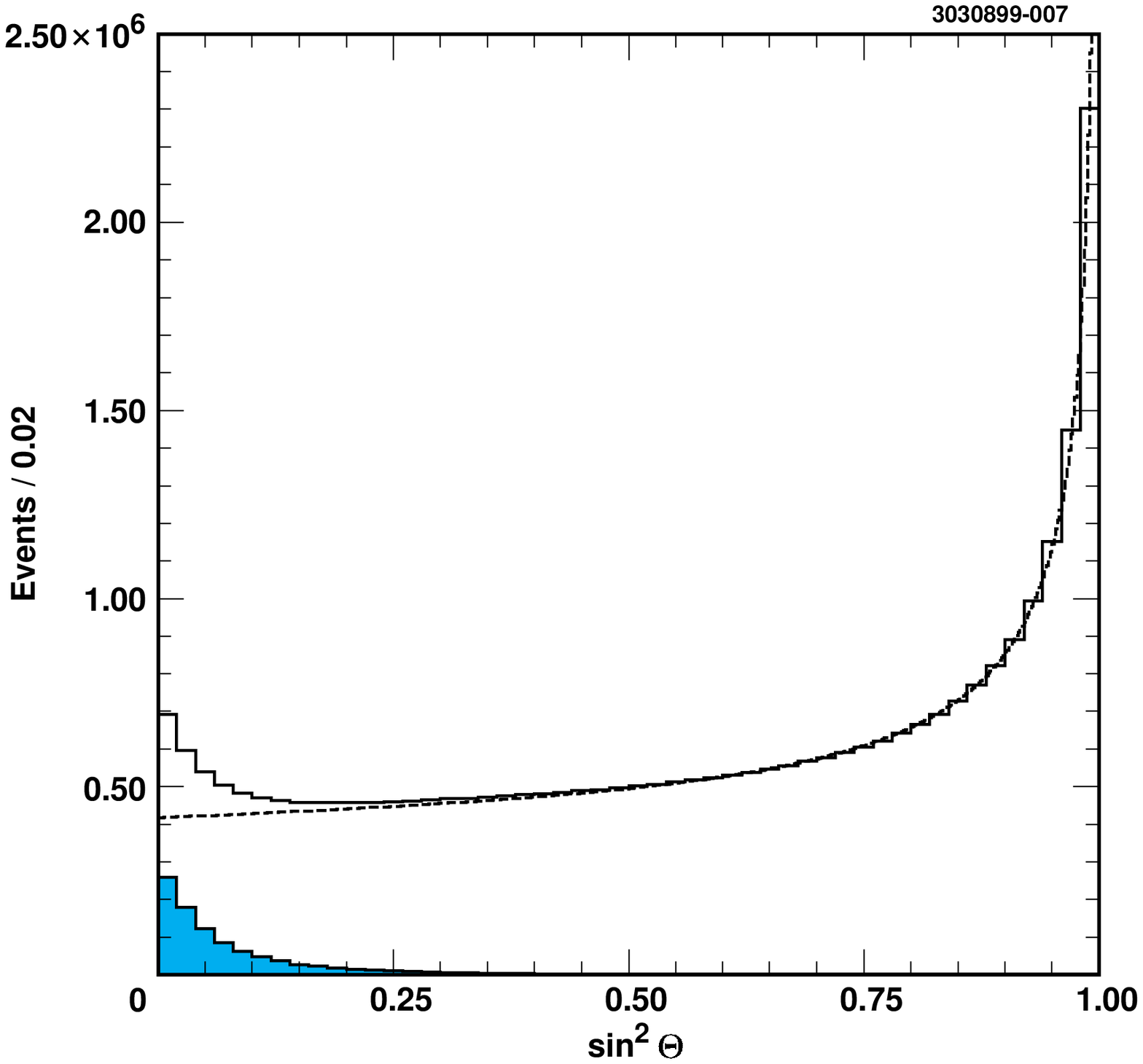}
\end{picture} \\
\caption{\label{fig:pion_sqrsine}
         \small $\sin^2 \theta$ distribution for candidate
                $\pi^-_{soft}$. Shown is the inclusive distribution (solid
histogram) overlaid with the background fit function (dashed) and the 
signal expected from $D^{*-}$ decays (shaded).}
\end{figure}

(d) The pion's trajectory must lie near the trajectory of the parent
charm quark, as expected for pions produced in
$D^{*-} \to \overline{D}^0 \pi^-_{soft}$.
Experimentally, this is checked using the variable sin$^2
\theta$, with $\theta$ the opening angle between 
the candidate soft pion
and the event thrust axis. Assuming that the thrust axis approximates
the original c\=c axis, true $\pi^-_{soft}$ should populate the region
$\sin^2 \theta \to 0$.  Figure~\ref{fig:pion_sqrsine}
displays the sin$^2 \theta$
distribution for candidates passing our event and track selection criteria.
The excess in the region $\sin^2 \theta$ near 0 constitutes our charm-tagged
sample.
The fit includes a signal contribution,
the shape of which is determined from Monte Carlo simulations, and a 
lower-order polynomial to fit the background. 
The technique for determining
the signal shape and background follows 
that of an earlier CLEO analysis\onlinecite{DKpiBR},
which used this method to measure the branching fraction 
${\cal B}(D^0\to K^-\pi^+)$.

\subsection{${\overline D^0}\to K^+\pi^-$ and $D^-\to K^+\pi^-\pi^-$ tags}
The ${\overline D}$-tagged analysis was performed 
independent of the $\pi^-_{soft}$ and
electron-tagged analyses.
For this latter analysis, we take advantage of 
improved track and particle reconstruction algorithms, 
which were unavailable when the 
$\pi^-_{soft}$
electron-tagged analyses were conducted. Also,
in order to compensate for the
intrinsically smaller efficiency of 
${\overline D^0}\to K^+\pi^-$ and $D^-\to K^+\pi^-\pi^-$ reconstruction,
we also use a three-fold larger data sample (13.1 fb$^{-1}$,
including the CLEO II.V 
data set\cite{CLEO-IIV}) for this analysis. Note
that, for the purpose of this analysis,
which does not utilize the precision vertexing afforded by the CLEO II.V
silicon vertex system,
the essential detector performance characteristics are 
the same as for the CLEO II data sample.
${\overline D^0}\to K^+\pi^-$ and $D^-\to K^+\pi^-\pi^-$ tags are
fully reconstructed from
kaon and pion candidates as follows:

(a) The kaon and
pion candidates must have impact parameters relative to the event vertex
less than 5 mm along the radial coordinate and no more than 5 cm along
the beam axis.

(b) Both the pion and kaon tracks must be consistent with their 
assumed  
particle identities at the level of 2.5 standard
deviations ($\sigma$), using the available
specific ionization and time-of-flight particle identification information.

(c) Both the pion and kaon must have momentum greater than 0.3 GeV/c.

(d) The fully reconstructed ${\overline D}$ meson tag
must have momentum greater than 2.3 GeV/c to eliminate 
B\=B backgrounds.

%%%%%%%%%%%%%%%%%%%%%%
%%%% Lambda Detection
%%%%%%%%%%%%%%%%%%%%%%

\section{Lambda Detection}
\label{sec:lambda}

\begin{figure}
\begin{picture}(180,220)
\includegraphics{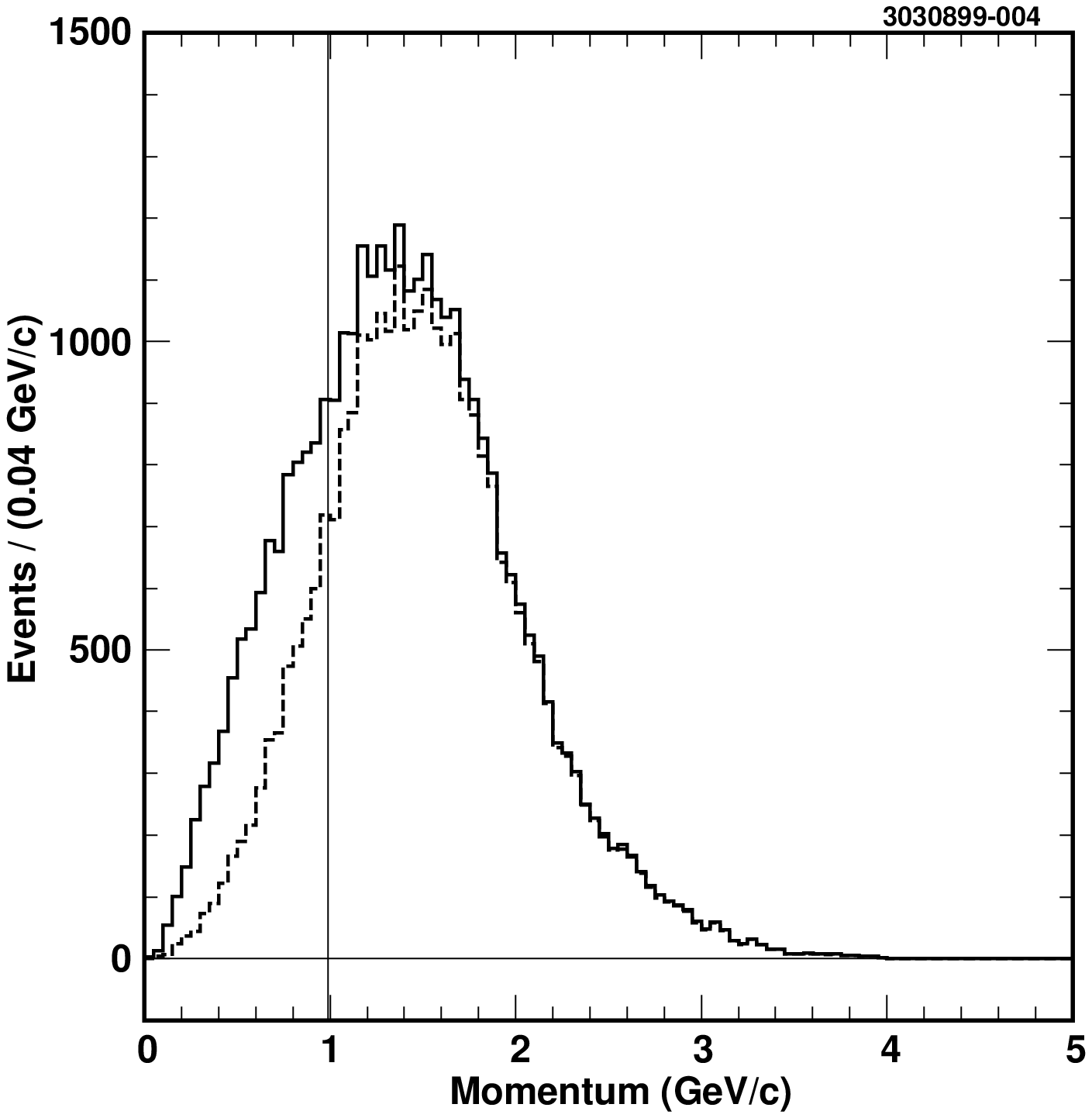}
\end{picture}
\begin{picture}(180,220)
\includegraphics{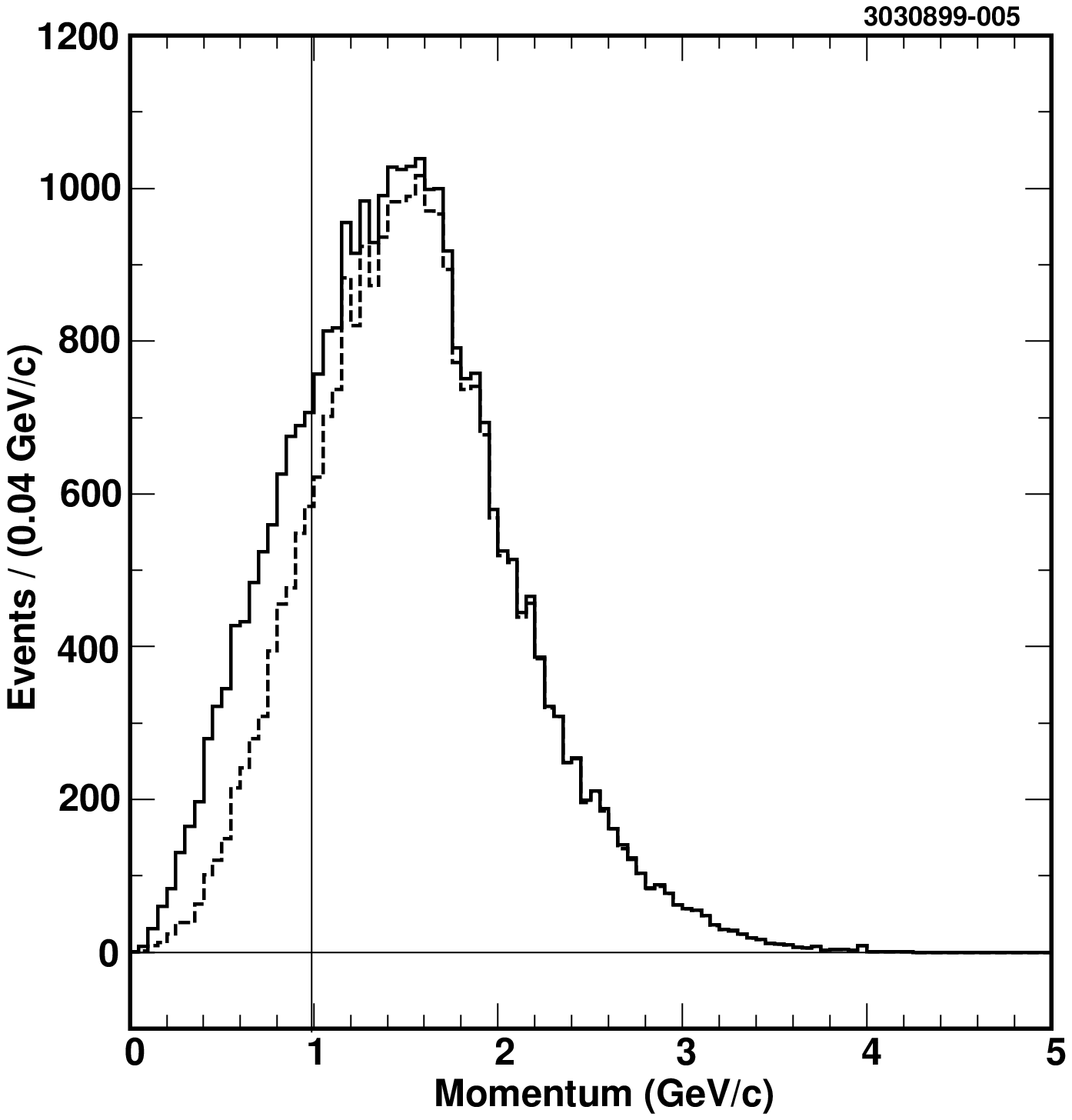}
\end{picture} \\
\caption{\label{fig:lam_from_lamc}
         \small JETSET 7.3 Monte Carlo simulations of
momentum spectra for $\Lambda$'s passing our tag, event, and 
Opposite Hemisphere/Opposite Sign
(OH/OS) requirements, for electron tags (left) and soft pion
tags (right).
Shown are
all $\Lambda$'s in c\=c events (solid) compared to
$\Lambda$'s which decay
from charmed baryons (dashed).
Our minimum momentum requirement 
($p_\Lambda>$1 GeV/c) is also indicated.}
\end{figure}

After finding a charm tag, we reconstruct $\Lambda \to p \pi$ in the 
hemisphere opposite the tag.
In addition to a 99\% particle identification 
probability requirement placed on both the
daughter proton and pion used in reconstructing the $\Lambda$,
we also require that
candidate $\Lambda$ particles have
momenta greater than 1 GeV/c and
that the lambda vertex be located at least 2 cm away from the $e^+e^-$
collision point in the radial direction.  
According to Monte Carlo simulations (Figure~\ref{fig:lam_from_lamc}),
imposing the minimum $\Lambda$ momentum
requirement (p$_\Lambda > 1.0$ GeV) in a charm-tagged event
passing our event selection requirements results in a $\Lambda$
sample which is $>$95\% pure $\Theta_c \to \Lambda X$, with
the remaining $\Lambda$'s due to light quark fragmentation.

%-%\begin{figure}
%-%\begin{picture}(200,250)
%-%\special{psfile=/home/ds/jrh/Paper/lam_from_lamc_elec.ps
%-%         voffset=-90 hoffset=-50
%-%         vscale=50 hscale=50
%-%         angle=0}
%-%\end{picture}
%-%\begin{picture}(200,250)
%-%\special{psfile=/home/ds/jrh/Paper/lam_from_lamc_pion.ps
%-%         voffset=-90 hoffset=0
%-%         vscale=50 hscale=50
%-%         angle=0}
%-%\end{picture} \\
%-%\caption{\label{fig:lam_from_lamc}
%-%         \small JETSET 7.3 Monte Carlo predictions for
%-%momentum spectra for $\Lambda$'s passing our tag, event, and 
%-%opposite hemisphere / opposite sign requirements. Shown are
%-%all $\Lambda$'s in c\=c events (solid) compared to
%-%$\Lambda$'s which decay
%-%from charmed baryons (dashed) in electron tagged (left) and
%-%soft pion tagged (right) events. Our minimum momentum cut 
%-%($p_\Lambda>$1 GeV/c) is also indicated.}
%-%\end{figure}

\section{Yields}
\label{sec:Signals}

\begin{figure}
\begin{picture}(200,250)
\includegraphics{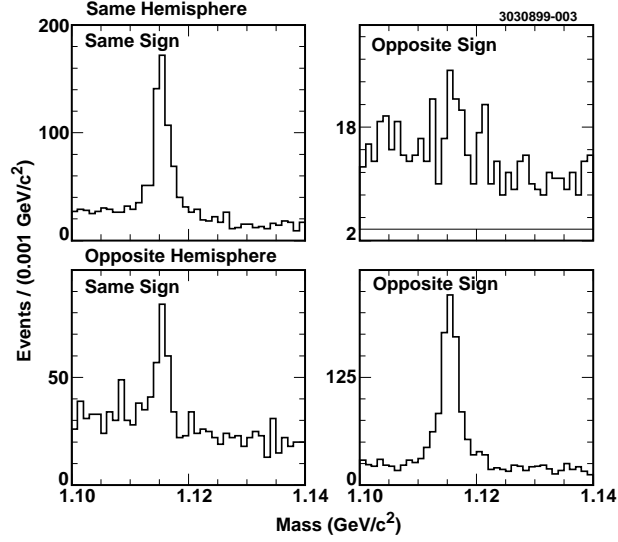}
\end{picture} \\
\caption{\label{fig:lam_corr_elec}
         \small Candidate $\Lambda$ particles from data divided into
         same/opposite hemisphere/sign correlations using the electron tag.
         Opposite Hemisphere / Opposite Sign $\Lambda$'s (lower right)
result predominantly from
         $\Lambda_c \to \Lambda X$; $\Lambda$-electron correlations
         can be seen from processes
such as $\Theta_c \to \Lambda e^- \nu_e X$ decays (same
         hemisphere / same sign), $\overline{c} \to \Lambda 
{\overline \Theta_c}, {\overline\Theta_c} \to e^-
         \nu_e X$ decays (same hemisphere / opposite sign) and $Xe^-\nu_e 
         \leftarrow \overline{c}c \to \Theta_c {\overline\Lambda} $ decays
         (opposite hemisphere / same sign).  From the known electron fake
	 rate,
         we conclude that fakes contribute from $\sim$10--40 events to each
         of the $\Lambda$ peaks.  Since the hemisphere correlation is not
         rigorous, we also note that
some of the ``wrong hemisphere'' correlation is due to
         leakage into the opposite hemisphere from real $\Theta_c \to
         \Lambda X$ or real semileptonic decays.}
\end{figure}

To extract our signal yields in the lepton-tagged sample, we plot the
proton-pion invariant mass for $\Lambda$ candidates in
electron-tagged events.
Figure~\ref{fig:lam_corr_elec} 
shows the candidate $\Lambda$ mass separated into
each of the four possible sign/hemisphere correlations.
Our candidate signal $\Lambda$'s are contained in
Figure~\ref{fig:lam_corr_elec}d (lower right).  
The number of signal $\Lambda$'s is extracted
by fitting a Gaussian $\Lambda$ signal function plus a 
second-order Chebyschev polynomial background.

%-%\begin{figure}
%-%\begin{picture}(200,250)
%-%\special{psfile=/home/ds/jrh/Paper/lam_corr_elec.ps
%-%         voffset=-90 hoffset=70
%-%         vscale=50 hscale=50
%-%         angle=0}
%-%\end{picture} \\
%-%\caption{\label{fig:lam_corr_elec}
%-%         \small Candidate $\Lambda$ particles from data broken into
%-%         same/opposite hemisphere/sign correlations using the electron tag.
%-%         While Opposite Hemisphere / Opposite Sign $\Lambda$'s (lower right)
%-%result predominantly from
%-%         $\Lambda_c \to \Lambda X$, $\Lambda$-electron correlations
%-%         can be seen from processes
%-%such as $\Theta_c \to \Lambda e \nu_e X$ decays (same
%-%         hemisphere / same sign), $\overline{c} \to \Lambda 
%-%{\overline \Theta_c}, {\overline\Theta_c} \to e
%-%         \nu_e X$ decays (same hemisphere / opposite sign) and $Xe\nu_e 
%-%         \leftarrow \overline{c}c \to \Theta_c {\overline\Lambda} $ decays
%-%         (opposite hemisphere / same sign).  From the known electron fake
%-%	 rate,
%-%         we conclude that fakes contribute approximately 20--40 events to ea%-%ch
%-%         of the $\Lambda$ peaks.  Since the hemisphere correlation is not
%-%         rigorous, we also note that
%-%some of the ``wrong hemisphere'' correlation is due to
%-%         leakage into the back hemisphere from real $\Theta_c \to
%-%         \Lambda X$ or real semileptonic decays.}
%-%\end{figure}

\begin{figure}
\begin{picture}(200,250)
\includegraphics{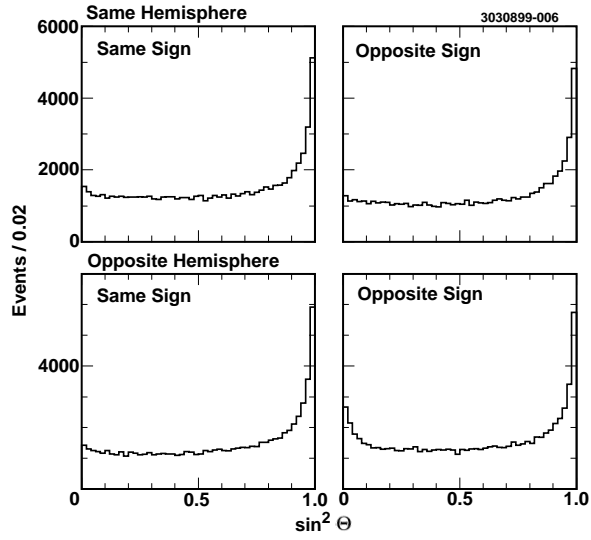}
\end{picture} \\
\caption{\label{fig:lam_sqrsine}
         \small Four possible $\Lambda/\pi^-_{soft}$ correlations,
after $\Lambda$ sideband subtraction (as described in text).}
\end{figure}

To determine the $\Lambda$ yield in $\pi^-_{soft}$ tagged events, we plot
the $\sin^2\theta$ of 
the $\pi$/thrust axis angle for each candidate $\Lambda$ found vs. the
candidate $p^+\pi^-$ mass. We then project the resulting histogram onto the
 $\sin^2 \theta$ axis and fit the peak at $\sin^2 \theta \to 0$ for the
case where the $p^+\pi^-$ invariant mass is in the $\Lambda$ region (signal)
versus the case where the $p^+\pi^-$ invariant mass is in the $\Lambda$
sidebands.
After performing a
sideband subtraction in $\Lambda$ mass of the two
$\sin^2\theta$ distributions, we obtain
Figure~\ref{fig:lam_sqrsine}.  A fit to the $\sin^2 \theta$ distribution
for all pion candidates (Figure~\ref{fig:pion_sqrsine}) 
determines our total number
of $\pi^-_{soft}$-tagged c\=c events (the denominator in our ratio). 

%-%\begin{figure}
%-%\begin{picture}(200,250)
%-%\special{psfile=/home/ds/jrh/Paper/lam_sqrsine.ps
%-%         voffset=-90 hoffset=70
%-%         vscale=50 hscale=50
%-%         angle=0}
%-%\end{picture} \\
%-%\caption{\label{fig:lam_sqrsine}
%-%         \small Four possible $\Lambda/\pi_{soft}$ correlations,
%-%after $\Lambda$ sideband subtraction (as described in text).}
%-%\end{figure}

Monte Carlo simulations indicate that both (i) the
fraction of non-$c\overline{c}$ tags and (ii) the fraction of 
candidate signal
$\Lambda$'s that do not originate from $\Theta_c$ decays but pass our
selection criteria are small ($<$3\%, see
Figure \ref{fig:lam_from_lamc}).  We test the 
overall accuracy of the Monte Carlo
by comparing
same-sign, opposite hemisphere correlation events
(``SS/OH'', i.e. opposite of the sign correlation expected for signal)      
in simulations compared to data.  We find that the ratio of
SS/OH electron-$\Lambda$ correlation
events to the number of ``right sign" (OS/OH) signal events is 
$0.19\pm0.05$ in Monte Carlo and $0.19\pm0.07$ in data.
The corresponding values for the
$\pi_{soft}^- - \Lambda$ correlations are 0.09 and 0.16,
respectively. Within statistics, the Monte Carlo reproduces the
``wrong sign'' (SS/OH) fractions observed in data.
%We note that the wrong-sign
%events are likely to result from processes which do not bear on our
%measurement. 
%(e.g., $c{\overline c}\to\Xi_c{\overline\Lambda}-{\overline c}-tag$).
Nevertheless,
we conservatively assign a relative
systematic error of 10\% (19\%/2) to reflect our confidence in
the simulations.
This value is entered in the final systematic errors table (Table IV) as
``Event generator mismodeling''. 

\begin{figure}
\begin{picture}(200,250)
\includegraphics{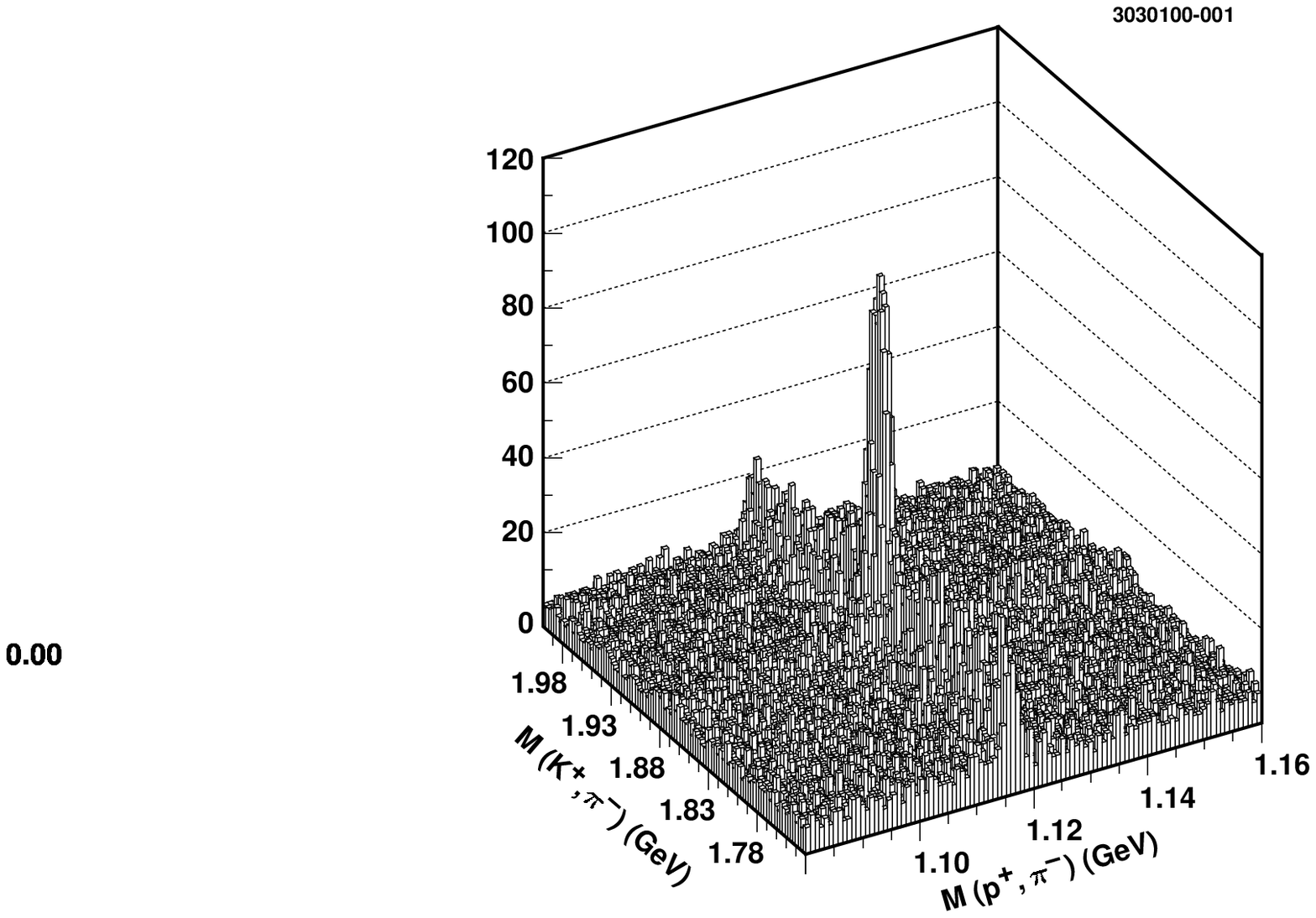}
\end{picture} \\
\caption{\label{fig:d0bar-lam-corr}
         \small Candidate
${\overline D^0}\to K^+\pi^-$ mass vs. opposite hemisphere
candidate $\Lambda\to p\pi^-$ mass.}
\end{figure}

\begin{figure}
\begin{picture}(200,250)
\includegraphics{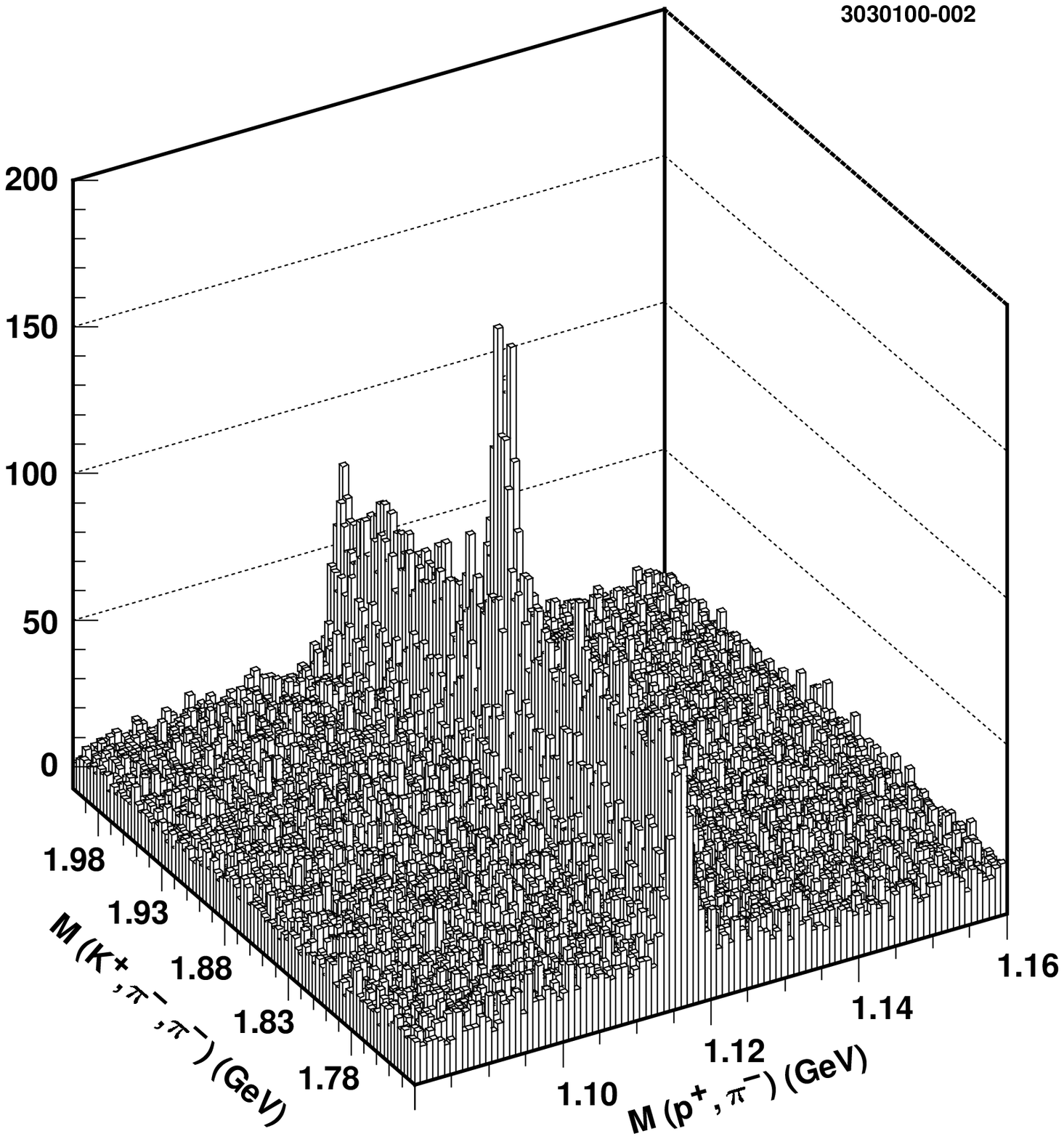}
\end{picture} \\

\vspace{1 cm}

\caption{\label{fig:dminus-lam-corr}
         \small Candidate
$D^-\to K^+\pi^-\pi^-$ mass vs. opposite hemisphere
candidate $\Lambda\to p\pi^-$ mass.}
\end{figure}

For the ${\overline D}$-tagged sample, the signal is extracted from a
two-dimensional plot of $M_{\Lambda_c}$ (the mass of
the $\Lambda_c$ candidate) vs. $M_{\overline D}$ (the mass of the 
${\overline D}$-tag candidate, either $M_{K^+\pi^-}$ or
$M_{K^+\pi^-\pi^-}$), as indicated in 
Figures \ref{fig:d0bar-lam-corr} and \ref{fig:dminus-lam-corr}.
Our signal comprises events which contain both a fully
reconstructed ${\overline D}$ and also a $\Lambda$ (``double-tags'').
The double-tag
signal yields are determined by a two-dimensional
sideband
subtraction technique, similar to that used to determine the signal
yield for the soft-pion tagged sample. Here, we subtract the scaled
$\Lambda$ yield in the ${\overline D}$ sideband region from the
$\Lambda$ yield in the ${\overline D}$ signal region. The resulting
excess, is, by definition, our double-tag signal.
As a check of the signal 
extraction, the yield for the
``wrong-sign'' double-tag signal (i.e., $D^0\to K^-\pi^+$ 
vs. $D^0\to K^-\pi^+$) is similarly
extracted using the same subtraction. 
In such events, the expected true
correlated signal should be negligible and non-zero only
through doubly Cabibbo suppressed decays\cite{DCSD}; 
in fact, we find $-15\pm 25$ events
from a two-dimensional $M(K^-\pi^+)$ vs. $M(K^-\pi^+)$ plot, and
$-115\pm157$ events from the two-dimensional
 $M(K^-\pi^+\pi^+)$ vs. $M(K^-\pi^+\pi^+)$ plot.

%%%%%%%%%%%%%%%%%%
%%%% Calculations
%%%%%%%%%%%%%%%%%%

\section{Calculations}
\label{sec:calculations}

The product branching fraction ${\cal B}(c \to
\Theta_c) \cdot {\cal B}(\Theta_c \to \Lambda X$) can be
derived from the fraction of times that an event
containing a charm tag in one hemisphere
contains a $\Lambda$ in the opposite hemisphere.
This, effectively, is the fraction of $\Lambda$
particles per c\=c event and should equal
the probability of a $c$ quark fragmenting to produce
a $\Theta_c$ multiplied by the probability
of the $\Theta_c$ to decay to a $\Lambda$ multiplied
by the efficiency for detection of a $\Lambda$ in our
charm-tagged event sample. In equation form, defining
${N(\Lambda)}\over{c\overline{c}}$ as the ratio of the number of
reconstructed $\Lambda$'s 
in tagged $c\overline{c}$ events to the total number
of $c\overline{c}$ event tags, we have:
  
\begin{equation}
   \frac{N(\Lambda)}{c\overline{c}} = {\cal B}(c \to \Theta_{c}) \cdot
                                      {\cal B}(\Theta_{c} \to \Lambda X)
                                      \cdot \epsilon_{\Lambda,tagged}
\end{equation}
for both data and Monte Carlo.
Assuming that the Monte Carlo simulation accurately reproduces
the efficiency for finding a $\Lambda$ in a tagged
event ($\epsilon_{\Lambda,tagged}$ in this equation),
the yield of non c\=c tags ($<$4\%), and 
the fraction of non-signal
$\Lambda-tag$
correlations ($<$5\%),
we can then calibrate our observed value of
$\Lambda$'s per c\=c in data to Monte Carlo:
\begin{equation}
   {\frac{N(\Lambda)}{c\overline{c}}^{Data} \over
    \frac{N(\Lambda)}{c\overline{c}}^{MC}} =
      {{{\cal B}(c \to \Theta_c X) \cdot {\cal B}(\Theta_c \to \Lambda X)}^{Data} \over
       {{\cal B}(c \to \Theta_c X) \cdot {\cal B}(\Theta_c \to \Lambda X)}^{MC}};
\end{equation}
the Monte Carlo values for ${\cal B}(c \to \Theta_c X)$
and ${\cal B}(\Theta_c \to \Lambda X)$ are 0.0667 and 0.369,
respectively. (A recent measurement by the ALEPH 
collaboration\cite{ALEPH99-ctolamc} at $\sqrt{s}$=90 GeV has 
determined ${\cal B}(c \to \Lambda_c X)=0.079\pm0.008\pm0.004\pm0.020$ 
[the last
systematic error represents the uncertainty in the $\Lambda_c\to pK^-\pi^+$
branching fraction],
although it is not clear how appropriate this value is
for $\sqrt{s}$=10 GeV.)
Note 
that the efficiency $\epsilon_{\Lambda,tagged}$ is
tag-dependent - due to geometric and momentum 
correlations from hemisphere to hemisphere,
we expect the highest efficiency for the ${\overline D}^0$
and $D^-$ tags, followed by soft pion tags and electron tags.

A summary of our yields 
and calculations for ${\cal B}$($c \to \Theta_c X$)
$\cdot$ ${\cal B}$($\Theta_c \to \Lambda X$) is presented
in Tables~\ref{table:lc2l} and \ref{table:lc2lcorrelation}.
Presented in those Tables are our raw yields, the number of true
electrons which do not tag c\=c events (`fake tags'), and the number
of $\Lambda$'s reconstructed in the opposite hemisphere for our
electron-tagged, soft pion-tagged, and ${\overline D}$-tagged samples. 
Backgrounds in the electron-tagged sample from B\=B and $\tau{\overline\tau}$
events are estimated from a large sample of Monte Carlo events,
using a CLEO event generator for B decays, and KORALB\cite{tauola} 
for $\tau{\overline\tau}$
decays. The electron background from $\gamma\gamma$ events is estimated
from the forward-backward excess of positrons versus electrons, compared to
the expectation from QED. 
Our yields correspond to 
${\cal B}_\Lambda$ = (1.62 $\pm$ 0.10)\%
using electrons to tag c\=c events, ${\cal B}_\Lambda$ = 
(1.53 $\pm$ 0.06)\% using $\pi^-_{soft}$ to tag c\=c events,
${\cal B}_\Lambda$ = (2.12 $\pm$ 0.09)\%
using ${\overline D^0}$ to tag c\=c events, and
${\cal B}_\Lambda$ = 
(2.09 $\pm$ 0.13)\% using $D^-$ to tag c\=c events (statistical errors
only).

%%%%%%%%%%%%%%%%%%%%%%%%%%%%%%%
%%%% Cross Checking Our Method
%%%%%%%%%%%%%%%%%%%%%%%%%%%%%%%

\section{Cross Checks}
\label{sec:cross_check}

We have conducted two cross-checks to verify the accuracy of our derived
result for ${\cal B}_\Lambda$. We emphasize that these are not measurements
in themselves (and therefore have no quoted systematic errors), but are
presented only to verify our ${\cal B}_\Lambda$
measurement.

\subsection{$D^0 \to K^- \pi^+$ decays.}
\label{sec:d02kpi}

As a first cross-check, we compare the data- versus Monte Carlo-derived
values for
the product branching
fraction:
${\cal B}(c\to D^0)\cdot 
{\cal B}(D^0 \to K^- \pi^+)$, using charm-tagging.
Since the branching
fraction for $D^0\to K^-\pi^+$ is known precisely, and since the fractional
uncertainty in 
${\cal B}(c\to D^0)$ 
is expected to be smaller than the corresponding
uncertainty in ${\cal B}(c\to\Lambda_c)$, we can compare
the value of ${\cal B}(c\to D^0)\cdot{\cal B}(D^0 \to K^-\pi^+)$ 
measured with charm-tagging in data versus Monte Carlo simulations and
thereby verify the method used in the $\Lambda$ measurement.
Using the same charm-tagged sample as before, we therefore search for
the decay $D^0\to K^-\pi^+$ (using the same
requirements mentioned before) opposite the tag
rather than $\Lambda\to p\pi^-$.
%If these requirements are met, we plot the invariant mass of the two 
%candidate $D^0$ daughter
%particles, as well as the $\sin^2 \theta$ distribution in the case of soft
%pion tagged events.  
As before, we perform a sideband subtraction to
determine the number of $D^0 \to K^- \pi^+$ decays in our
$\pi^-_{soft}$
charm tagged sample.  We thus use the same equation as with our $\Lambda$
analysis, only modified for the $D^0 \to K^- \pi^+$ decay mode:
\begin{equation}
   \frac{N(K^-\pi^+)}{c\overline{c}} = {\cal B}(c \to D^0) \cdot
                                       {\cal B}(D^0 \to K^-\pi^+)
                                       \cdot \epsilon_{D^0\to K^-\pi^+}
\end{equation}

Again, assuming that the Monte Carlo accurately reproduces
the efficiency for finding a $D^0$ decay in a tagged event,
we calibrate our observed value of $D^0$'s per c\=c in data
to Monte Carlo:
\begin{equation}
   {\frac{N(K\pi)}{c\overline{c}}^{Data} \over
    \frac{N(K\pi)}{c\overline{c}}^{MC}} =
	 {{{\cal B}(c \to D^0) \cdot {\cal B}(D^0 \to K\pi)}^{Data} \over
	  {{\cal B}(c \to D^0) \cdot {\cal B}(D^0 \to K\pi)}^{MC}}
\end{equation}
%where the Monte Carlo values for ${\cal B}(c \to D^0)$
%and ${\cal B}(D^0 \to K^-\pi^+)$ are 0.53 and 0.0391,
%respectively.  
%This measurement is then converted into
%a value for ${\cal B}(D^0 \to K^-\pi^+)$ by inserting the Monte
%Carlo estimated value for ${\cal B}(c \to D^0)$.
The results of our $D^0$ cross-check are presented in
Table~\ref{table:d02kpi}. 
The Monte Carlo adequately reproduces the 
$D^0\to K^-\pi^+$ yield per ${\overline c}$-tag.
Based on the consistency between these values
and the known $D^0 \to K^- \pi^+$
branching fraction, a scale factor is applied to the data
and a systematic error is added which reflects
only the statistical precision of this cross check.
In all cases, the scale factor (Table \ref{table:d02kpi}) is
consistent with unity.

\subsection{$\Lambda_c \to \Lambda e \nu_e$ decays.}
\label{sec:lc2lenu}

A second cross-check is afforded by our $\Lambda$-electron correlation
sample. We note that the same-hemisphere, same-sign events are
expected to be dominated by $\Lambda_c\to\Lambda e^+\nu_e$ decays. Since
the $\Lambda_c\to\Lambda e\nu_e$ branching fraction has been measured, 
we can use the relative ratio of the same-hemisphere, same-sign 
$\Lambda$-electron events, compared to the opposite-hemisphere, opposite-sign
events to estimate the $\Lambda_c\to\Lambda$X branching fraction.
This estimate is ``internally normalizing''; i.e., we do not need to 
measure the fraction of our total charm tags which contain $\Lambda$'s.
We can relate the
branching fractions ${\cal B}(\Lambda_c\to\Lambda e\nu_e)$ and
${\cal B}(\Lambda_c\to\Lambda X)$ (and their corresponding
efficiencies $\epsilon$) to the number of observed
same-hemisphere, same-sign events ($N_{SH/SS}$), the number of
observed opposite hemisphere, opposite sign events $N_{OH/OS}$,
and their production fractions in c\=c events. Without an
explicit fake electron subtraction to the observed yields, we have:

\message{MISSING SIGNS/NOTATION CONFUSING HERE}
$${N_{SH/SS}\over N_{OH/OS}}\approx {{\cal B}(c\to\Lambda_c)\cdot
{\cal B}({\Lambda_c\to\Lambda e\nu_e})
\cdot \epsilon(\Lambda_c\to\Lambda e\nu_e)
\over {\cal B}(c\to\Lambda_c)\cdot 
{\cal B}({\overline c\to eX})\cdot
{\cal B}({\Lambda_c\to\Lambda X})\cdot(\epsilon(\Lambda_c\to\Lambda X)
({\overline c}\to eX))}$$

Note that the efficiency in the numerator of this equation refers to the
correlated efficiency of having both the $\Lambda$ and the electron in
$\Lambda_c\to\Lambda e^+\nu_e$ pass all our selection criteria
($\epsilon(\Lambda_c\to\Lambda e\nu_e)=0.023\pm0.002$); the efficiency in
the denominator refers to the efficiency for having a $\Lambda$ from
a $\Lambda_c$ decay pass our selection
criteria in one hemisphere, and also an electron from
a generic charm decay pass our selection requirements 
in the opposite hemisphere 
($\epsilon(\Lambda_c\to\Lambda X)({\overline c}\to eX)=0.043\pm0.002$). 
The efficiency is lower in the numerator
due to the presence of momentum correlations between the $\Lambda$ and
the electron, resulting in a reduced efficiency for both
particles to simultaneously pass the minimum momentum requirement $p>$1 GeV/c.
The value for ${\cal B}(\overline c\to eX)$
(0.091$\pm$0.008)
is taken from data at $\sqrt{s}$=10 GeV\cite{pdg98}. 
Using the current Particle Data Group\cite{pdg98}
value for ${\cal B}(\Lambda_c\to\Lambda e\nu_e)=0.021\pm0.006$ and
our measured values for $N_{SH/SS}$ ($445\pm26$) and
$N_{OH/OS}$ ($743\pm32$), we obtain an inferred value of
${\cal B}(\Lambda_c\to\Lambda X)\sim 0.23\pm0.07$, where the error is
statistical 
only. This is consistent with our measured product branching fraction for
${\cal B}_\Lambda = {\cal B}(c \to \Theta_c)\cdot {\cal B}(\Theta_c \to \Lambda
X$) if 
${\cal B}(c\to\Theta_c$)=6.67\%, and assuming that
${{{\cal B}(c\to\Lambda_c)}\over{{\cal B}(c\to\Theta_c)}}\approx 
1.0$.%\footnote{The JETSET 7.3 value is 0.878 $\pm$ 0.003.}.

\section{Systematic Errors}
\label{sec:systematics}
In order to determine additional systematic errors, we varied each of our
individual particle and event requirements and noted the corresponding
variation in our derived values for ${\cal B}(c \to\Theta_c)\cdot 
{\cal B}(\Theta_c \to \Lambda X$). Typical variations were 
of order $\sim$20\%.  
Using
this approach,
we summarize our systematic dependencies in Table~\ref{table:systematics}.
The default values of, e.g., our kinematic cuts are defined, as well as the
variation used to assess systematic dependencies. 
For the pion and the electron tags, 
our largest systematic errors are due to uncertainties
in the Monte Carlo event generation modeling, as determined using the
``wrong-sign'' yields. 
For the ${\overline D}$-tags, among the largest errors
are the errors associated with signal extraction - this is assessed by
determining the difference in the calculated final result when the signal 
and sideband regions are varied from their default values by $\pm$30\%.
We also take the r.m.s. spread in the values for ${\cal B}_\Lambda$ 
obtained with the four tags (14\%) as an additional systematic error,
reflecting the differences in the lepton-tagged vs. 
$\pi^-_{soft}$-tagged vs.
D-tagged samples. This
variance is the dominant contributor to our overall quoted systematic error.

%%%%%%%%%%%%%
%%%% Summary
%%%%%%%%%%%%%

\section{Summary and Discussion}
\label{sec:summary}

Using four different $e^+e^-\to c{\overline c}$ tags,
we measure
the product branching fraction ${\cal B}_\Lambda$ = 
${\cal B}(c \to \Theta_c X)
\cdot {\cal B}(\Theta_c \to \Lambda X$):
\begin{center}
$(1.62 \pm 0.10 \pm 0.32)\%~~~~(electron~tags)$ \\
$(1.53 \pm 0.06 \pm 0.30)\%~~~~(soft~pion~tags)$ \\
$(2.12 \pm 0.09 \pm 0.30)\%~~~~({\overline D^0}~tags)$ \\
$(2.09 \pm 0.13 \pm 0.42)\%~~~~(D^-~tags);$
\end{center}

\noindent
these results sum over the charmed baryons $\Theta_c$ produced at 
$\sqrt{s}$=10 GeV. Separating common from independent systematic errors,
and weighting each result by the quadrature
sum of its statistical error plus independent 
systematic error,
we combine
these four numbers to obtain a weighted product branching fraction:
	\[
		{\cal B}_\Lambda = (1.87 \pm 0.03 \pm 0.33)\%
	\]

\begin{figure}
\begin{picture}(200,250)
\includegraphics{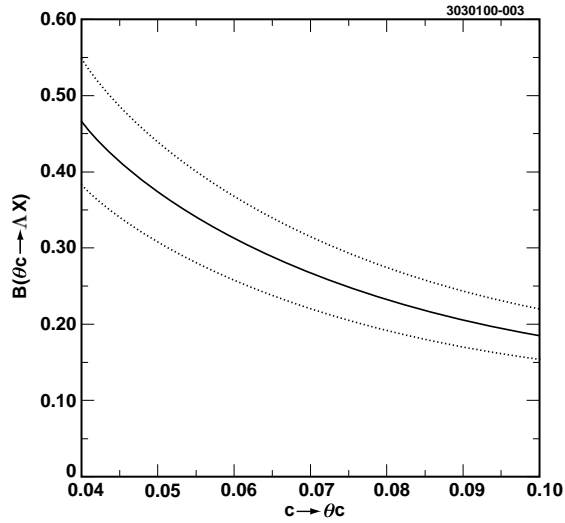}
\end{picture} \\
\caption{\label{fig:lamcontour}
         \small Contour of
${\cal B}(\Theta_c\to\Lambda X)$ vs. $c\to\Theta_c$
implied by our result. Contours corresponding to $\pm 1\sigma$ excursions
from our central value are also shown (dotted).
The Monte Carlo value for $B_\Lambda$ 
(${\cal B}(c \to \Theta_c) \cdot{\cal B}(\Theta_c \to \Lambda X)$ is 0.0246, 
and the Monte Carlo expectation for $c\to\Theta_c$ 
is 0.0667.}
\end{figure}

In obtaining this result, we have not corrected for the statistical overlap
between the four tag samples. Correcting for this would tend to slightly
reduce the overall quoted statistical error.
We can convert this result into a contour in the plane:
${\cal B}(\Theta_c\to\Lambda X)$ vs. ${\cal B}(c\to\Theta_c)$,
%\cdot{{{\calB}(c\to\Lambda_c)}\over{{\cal B}(c\to\Theta_c)}}$, 
as shown in Figure~\ref{fig:lamcontour}.
%-%\begin{figure}
%-%\begin{picture}(200,250)
%-%\special{psfile=lamclamx_contour.ps
%-%         voffset=-90 hoffset=70
%-%         vscale=50 hscale=50
%-%         angle=0}
%-%\end{picture} \\
%-%\caption{\label{fig:lamcontour}
%-%         \small Contour of
%-%${\cal B}(\Lambda_c\to\Lambda X)$
%-%vs. $c\to\Theta_c\cdot{\Lambda_c\over\Theta_c}$ 
%-%implied by our result.}
%-%\end{figure}
Using the Monte Carlo value for ${\cal B}$($c \to \Theta_c X$)
of 6.67\% (this is consistent with the tabulated
product cross-section
$(e^+e^-\to c{\overline c})\cdot{\cal B}(c\to\Lambda_c)\cdot{\cal
B}(\Lambda_c\to pK^-\pi^+)$ using ${\cal B}(\Lambda_c\to pK^-\pi^+)$=5.0\%),
and taking the results from our four tags,
we can infer a weighted average value:
	\[
		{\cal B}(\Theta_c \to \Lambda X) = (28 \pm 1 \pm 5)\%
	\]
It is important to note that this measurement is independent of the
$\Lambda_c \to pK^-\pi^+$ normalization but is dependent on the Monte
Carlo estimated value for ${\cal B}(c \to \Theta_c)$. 
This measurement is the first of its kind at $\sqrt{s}$ = 10 GeV. 

In the simplest picture, a charmed baryon such as a $\Lambda_c$ decays 
weakly through
external W-emission. Neglecting fragmentation at the lower vertex, this
%(Figure~\ref{fig:spec_model}a), 
produces either a $\Sigma^0$
or, if isospin does not change, a $\Lambda$.  Since all $\Sigma^0$'s decay
into $\Lambda$, we therefore expect that $\Lambda_c \to \Lambda X \approx
100\%$ if the external spectator diagram dominates.  This simple-minded 
prediction
is expected to be
obeyed in semileptonic decays; i.e., 
${\cal B}(\Lambda_c \to \Lambda l \nu)/{\cal B}(\Lambda_c \to X l
\nu) \to
1$. Present data, however, give a value of approximately 50\% for this ratio,
albeit with large errors\onlinecite{pdg98}.  

External and internal W-emission,
as well as W-exchange %(Figure~\ref{fig:spec_model}b) 
can lead to $NKX$
final states ($\Lambda_c\to pK^0_{\rm s}$, e.g.). 
The fact that the 
$\Lambda_c$ lifetime is only half that of the $D^0$ meson 
suggests that internal W-emission and W-exchange processes
may comprise a large fraction
of the total $\Lambda_c$ width.
Although internal W-emission may be suppressed in decays of charmed
mesons due to the color-matching requirement (which would predict
${\cal B}(D \to W_{int} X)/{\cal B}(D \to W_{ext} X) = 1/9$),
the larger number of degrees of freedom
in baryon decays may mitigate this suppression, leading to a 
potentially large
fraction of $pKX$ final states.
%It is also possible that other diagrams, such as
%W-exchange, may be 
%substantial in charmed baryon decays. 
In the case of the $\Lambda_c$, W-exchange
decays can produce either $\Lambda$'s or $NK$ in the final state,
depending on the quark configuration. The naive
expectation that the absence of exchange diagrams in $\Xi_c^+$ decays
will lead to a longer lifetime for $\Xi_c^+$ compared to 
$\Xi_c^0$ and $\Lambda_c$ is consistent with current experimental
data.
% (although this argument seems to not hold in the case of the $\Omega_c$).
%Observation of
%the decay $\Xi_c^0\to\Omega^- K^+$ provided compelling evidence for
%the presence of exchange diagrams in charmed baryon decays.

The current world average for ${\cal B}(\Lambda_c \to \Lambda X)$ (35 $\pm$
11)\% 
\onlinecite{pdg98} is consistent with the 
notion that the simple-minded external
W-emission picture does not saturate $\Lambda_c$ decays.
The value for ${\cal B}(\Lambda_c \to \Lambda X)$ therefore has implications
for the external versus internal spectator fractions in charmed baryon decay.
Our results are therefore qualitatively consistent with a possibly substantial
internal spectator contribution to charmed baryon decay.

Exclusive $\Lambda_c \to \Lambda X$ channels have also
been measured; normalized to an
estimate of ${\cal B}(\Lambda_c \to pK^-\pi^+) = (5.0 \pm 1.3)\%$
\onlinecite{pdg98,cleo_pkpi00}, the sum of the observed exclusive modes 
account for
the bulk
of the presently tabulated inclusive $\Lambda_c \to
\Lambda X$ rate
(${\Sigma {\cal B}(\Lambda_c\to\Lambda+X)_{exclusive}\over
{\cal B}(\Lambda_c\to pK^-\pi^+)}\sim$5, where the sum includes a contribution
of ${\Sigma{\cal B}(\Lambda_c\to\Sigma^0+X)_{exclusive}\over
{\cal B}(\Lambda_c\to pK^-\pi^+)}\sim$1.5).
%of ${\cal B}(\Lambda_c\to\Lambda X$)=$35\pm11$\% \onlinecite{pdg98}.
We note that the 
difference between our inferred value for ${\cal B}(\Theta_c \to
\Lambda X)$ 
and the sum of the exclusive $\Lambda_c$ modes to $\Lambda$'s\onlinecite{pdg98}
suggests that
most of the inclusive $\Lambda_c\to\Lambda$X rate has been
accounted for.

If charmed baryons produced in $e^+e^-$ events are predominantly
$\Lambda_c$'s,
and if the JETSET expectation for $f(c \to \Lambda_c)$ is accurate, then our
results are in agreement with the current world average for
${\cal B}(\Lambda_c \to \Lambda X)$.  
We note that the methodology of this analysis
differs substantially from the previous CLEO 
analysis\cite{cleo92}, which relied on a model
of charmed baryon production in B-decay to derive ${\cal B}(\Lambda_c \to
\Lambda X)$. 

Naively, one might expect fragmentation and decay of charmed baryons to
be similar to bottom baryons.
Using vertex tagging techniques, the OPAL collaboration
has determined ${\cal B}(b \to \Theta_b){\cal B}(\Theta_b
\to \Lambda X) = (3.50 \pm 0.32 \pm0.35)\%$ \onlinecite{opal98}. 
Perhaps the simplest way to reconcile the two numbers is to assume
({\it ad hoc}) that ${\cal B}(b\to\Theta_b)$ at $\sqrt{s}\sim$90 GeV is
approximately twice as large as
${\cal B}(c\to\Theta_c)$ at $\sqrt{s}$=10 GeV,
and that
${\cal B}(\Lambda_b\to\Lambda_c X)\approx$1.0.
However, the fact that the
$\Lambda_b$ lifetime is only 2/3 that of the
B-mesons \onlinecite{pdg98}, coupled with
the fact that $\Lambda_b$ has already been observed
through $\Lambda_b\to\psi\Lambda$ imply
%suggests that modes such as
%$\Lambda_b\to Dp$ may contribute substantially to $\Lambda_b$ decay. 
that
${\cal B}(\Lambda_b\to\Lambda_c X)<1$. 
Correspondingly, 
we expect an
enhancement of ${\cal B}(b\to\Theta_b)$ at OPAL
relative to ${\cal B}(c\to\Theta_c)$ at CLEO.

Finally, we stress that our final central value for
${\cal B}_\Lambda$
averages over the specific mix of
charm-tags that we use in this analysis.
The composition of our
${\overline D}$-meson tags will not be the same as the composition
of our electron tags, insofar as the lepton-tagged
sample represents a weighted
sum of ${\overline\Theta_c}\to l^-X$, $D_s^-\to l^-X$, 
${\overline D^0}\to l^-X$ and $D^-\to l^-X$. 
Our quoted final result can be
general only if the two hemispheres in an
$e^+e^-\to c{\overline c}$ event fragment independently.
Although not yet measured,
it is possible that there may be correlated ${\overline\Theta_c}\Theta_c$ 
production, in which case the likelihood of observing $\Theta_c\to\Lambda X$
would be larger for ${\overline\Theta_c}$ tags than for ${\overline D}$ tags,
and the assumption of independent fragmentation would be invalid.
A study of correlated ${\overline\Theta_c}\Theta_c$ production, presently
in progress, will be the subject of a forthcoming publication.

%%%%%%%%%%%%%%%%%%%%%%
%%%% Acknowledgements
%%%%%%%%%%%%%%%%%%%%%%

\acknowledgments
\label{sec:acknowledgments}

We gratefully acknowledge the effort of the CESR staff in providing us with
excellent luminosity and running conditions.
J.R. Patterson and I.P.J. Shipsey thank the NYI program of the NSF, 
M. Selen thanks the PFF program of the NSF, 
M. Selen and H. Yamamoto thank the OJI program of DOE, 
J.R. Patterson, K. Honscheid, M. Selen and V. Sharma 
thank the A.P. Sloan Foundation, 
M. Selen and V. Sharma thank the Research Corporation, 
F. Blanc thanks the Swiss National Science Foundation, 
and H. Schwarthoff and E. von Toerne thank 
the Alexander von Humboldt Stiftung for support.  
This work was supported by the National Science Foundation, the
U.S. Department of Energy, and the Natural Sciences and Engineering Research 
Council of Canada.

%%%%%%%%%%%%%%%%
%%%% References
%%%%%%%%%%%%%%%%

%%%%%%%%%%%%%
%%%% Figures
%%%%%%%%%%%%%

\newpage

%%%%%%%%%%%%
%%%% Tables
%%%%%%%%%%%%

\newpage

\begin{table}
\caption{\label{table:lc2l}
         Results 
for ${\cal B}(c \to \Lambda_c) \cdot {\cal B}(\Lambda_c \to \Lambda X)$
         using electrons, $\pi^-_{soft}$, and ${\overline D}$ 
to tag c\=c events. }%
\begin{tabular}{lc|cccc}
   Tag Particle & & Electron  & $\pi$ & ${\overline D^0}$ & $D^-$ \\ \hline
   Total tags  & & 177700$\pm$422 & 1060870$\pm$3942 & 325200$\pm$1040 &
256730$\pm$1180      \\
   Fake tags & $\gamma\gamma$           & 4665    $\pm$ 98
                                        &                        \\
             & $\tau\tau$               & 9429    $\pm$ 97
                                        &                        \\
             & B\=B                     & 15037   $\pm$ 123
                                        &                        \\
   Total Corrected Tags                        & & 148569  $\pm$ 460
                                        & 1060870 $\pm$ 3942 &
 325200$\pm$1040 & 256730$\pm$1180        \\
   $\Lambda$'s                        & & 743     $\pm$ 32 & 
5964$\pm$203 &
2247$\pm$75 &
2031$\pm$102 \\
   $\frac{N(\Lambda)}{c\overline{c}}$ ($\times 10^{-4}$)& & 50$\pm$2
                                        & 56$\pm$2 & 69$\pm$2 & 79$\pm$4    \\ 
\hline 
Monte Carlo {\bf $\frac{N(\Lambda)}{c\overline{c}}$} 
($\times 10^{-4}$) & & 76 $\pm$ 3.3
                                    & 90 $\pm$ 2.2 & $80\pm1.9$ & $93\pm3.7$ \\
DATA $B_\Lambda$ 
%${\cal B}(c \to \Theta_c) \cdot {\cal B}(\Theta_c \to \Lambda X) $
& & 0.0162 $\pm$ 0.0010
                                           & 0.0153 $\pm$ 0.0006
& $0.0212\pm0.0009$ & $0.0209\pm0.0013$ \\
\end{tabular}
\end{table}

\begin{table}
\caption{\label{table:lc2lcorrelation}
         Tabulated number of $\Lambda$'s detected for both signal
         and non-signal correlations followed by
         $\Lambda/c\overline{c}$. For $\pi^-_{soft}$ and $e^-$, a
data sample corresponding to $\sim$5 million c\=c events were used;
${\overline D^0}$ and $D^-$ tags are derived from a dataset of
$\sim$15 Million c\=c events. In both cases, the Monte Carlo
simulated data sample is of comparable, or slightly larger size.
Shown are opposite
hemisphere (OH), same hemisphere (SH), opposite sign (OS) and same
sign (SS) correlations.}
\begin{center}
\begin{small}
\begin{tabular}{c|cccc}
 Correlation                        & Electron  & $\pi$ & ${\overline D^0}$ &
$D^-$            \\ \hline
 {\bf Data}                         &           &                   \\
 SH, SS         & 445 $\pm$ 26 & 1039 $\pm$ 170 & 252$\pm$33 & 220$\pm$34\\
 SH, OS     & 43  $\pm$ 13 & 484  $\pm$ 186 & 14$\pm$24 & 12$\pm$38 \\
 OH, SS     & 142 $\pm$ 19 & 914  $\pm$ 356 & 223$\pm$76 & 233$\pm$91\\
 OH, OS & 743 $\pm$ 32 & 5964 $\pm$ 203 & 
2247$\pm$75 & 2031$\pm$102\\
                                                                    \\
 $\frac{\Lambda}{c\overline{c}}$ 
SH, SS ($\times 10^{-4}$)        & 32.3 $\pm$ 1.9
                         & 9.8 $\pm$ 1.6 & 7.8$\pm$1.0 & 8.6$\pm$1.3   \\
 $\frac{\Lambda}{c\overline{c}}$ 
SH, OS ($\times 10^{-4}$)    & 3.1 $\pm$ 0.9
                         & 4.6 $\pm$ 1.8 & $0.4\pm0.6$ & 0.4$\pm0.8$       \\
 $\frac{\Lambda}{c\overline{c}}$ 
OH, SS ($\times 10^{-4}$)    & 10.3 $\pm$ 1.4
                     & 9  $\pm$ 3 & 7.2$\pm$2.3  & 9.1$\pm$3.6 \\ \hline
{\bf $\frac{\Lambda}{c\overline{c}}$ 
OH, OS} ($\times 10^{-4}$) & 50$\pm$2
                         & 56$\pm$1.9  & $69\pm2$ & $79\pm4$ 
\\ \hline
                                                                    \\
 {\bf Monte Carlo}                  &                 &                 \\
 SH, SS         & 573$\pm$29   & 5469$\pm$363 & 182$\pm$52 & 179$\pm$65 \\
 SH, OS     & 62   $\pm$ 13   & 1448  $\pm$ 381 &  44$\pm$23 & 4$\pm$46 \\
 OH, SS     & 264  $\pm$ 24   & 2251  $\pm$ 409 & 285$\pm$104 & 114$\pm$62 \\
 OH, OS & 1352 $\pm$ 61   & 26413 $\pm$ 643 
& $2073\pm65$ & $1514\pm85$ \\
                                                                    \\
 $\frac{\Lambda}{c\overline{c}}$ 
SH, SS ($\times 10^{-4}$)        & 32 $\pm$ 2
                          & 19 $\pm$ 2  & $7\pm2$ & $11.2\pm4$       \\
 $\frac{\Lambda}{c\overline{c}}$ 
SH, OS ($\times 10^{-4}$)    & 4 $\pm$ 1
                                    & 5 $\pm$ 1 & $2\pm1$  &$3\pm2$      \\
 $\frac{\Lambda}{c\overline{c}}$ 
OH, SS ($\times 10^{-4}$)    & 15 $\pm$ 2
                                    & 8  $\pm$ 2  & $11\pm5$ & $7\pm4$        \\ \hline
Monte Carlo {\bf $\frac{\Lambda}{c\overline{c}}$
OH, OS} ($\times 10^{-4}$) & 76 $\pm$ 3
                                    & 90 $\pm$ 2 & $80\pm2$ & $93\pm4$
\\ \hline
\end{tabular}
\end{small}
\end{center}
\end{table}

\newpage 

\begin{table}
\caption{\label{table:d02kpi}
         Results of analyzing the ${\cal B}(D^0 \to K^- \pi^+)$ cross check.}
\begin{tabular}{lc|cccc}
   Tag Particle                       & & Electron  & $\pi$ & ${\overline D^0}$
& $D^-$       \\ \hline
   Total tags                         & & 177700  $\pm$ 422 
& 1060870 $\pm$ 3942 & 325200$\pm$1040 &
256730$\pm$1180 \\
   Fake tags & $\gamma\gamma$           & 4665    $\pm$ 68
                                        &                       \\
             & $\tau\tau$               & 9429    $\pm$ 97
                                        &                       \\
             & B\=B                     & 15037   $\pm$ 123
                                        &                       \\
   Actual tags                        & & 148569  $\pm$ 460
                                        & 1060870 $\pm$ 3942 & 325200$\pm$1040
 & 256730$\pm$1180    \\
  $D^0$'s                             & & 1154    $\pm$ 41
                                        & 10029   $\pm$ 213 & $1965\pm61$ &
$1244\pm144$       \\
Data  $\frac{N(D^0)}{c\overline{c}}$ ($\times 10^{-4}$)     & & 78$\pm$3
                                        & 94$\pm$2 &
$121\pm4$ & $123\pm9$   \\
Monte Carlo  $\frac{N(D^0)}{c\overline{c}}$ ($\times 10^{-4}$) & & 
$80\pm3$
                                        & $96\pm2$ &
$119\pm4$ & $113\pm11$    \\
%  ${\cal B}(D^0 \rightarrow K^- \pi^+)$ & & 0.038 $\pm$ 0.0020 
%                                          & 0.039 $\pm$ 0.0012
Scale factor && $1.03\pm0.05$ & $1.02\pm0.03$ & $0.98\pm0.06$ & $0.92\pm0.12$
\\
\end{tabular}
\end{table}

%0.0384
%0.0386

\begin{table}
\caption{\label{table:systematics}
         Summary of systematic errors for both electron, soft pion, 
${\overline D^0}$ and $D^-$-tagged events. 
Correlated systematic errors are indicated with
an asterisk ($^*$).}
%\tabcolsep[]
\begin{tabular}{lcccc}
  Systematic   (Variation/{\bf Default})    & $e^-$  & $\pi^-_{soft}$ Tag  &
${\overline D^0}$ & $D^-$ \\ \hline
  \multicolumn{3}{l}{\em Event Requirements}                               \\
  Minimum event energy:    0.88--1.32$E_{beam}$ 
(${\bf 1.1E_{beam}}$)        & 2\%         & 3\%   & 2\% & 2\%    \\
  Minimum number of charged tracks: 3--7 (4)
& 7\%         & 3\%    & 2\% & 2\%   \\ \\
  \multicolumn{3}{l}{\em Tag Requirements}                                 \\
  Tag momentum:   0.8--1.2$\times$ default & 8\% & 3\% & 4\%  & 4\%  \\
  Radially close to event vertex: 0.8--1.2$\times$ default & 2\% & 3\% & & \\
  Tag DOCA along beam axis:     4--6 cm ({\bf 5 cm}) & 2\%    & 3\% & &      \\
  Probability of being a pion:  2.4$\sigma-3.6\sigma$  (${\bf 3\sigma}$) &    
           & 3\%     & &  \\
Minimum Electron Probability: (Ln$({\tt L}_e)\ge$7 
(Ln$({\tt L}_e)\ge$5)  & 9\% & & \\
  Event Sphericity Cut: 0.25$\le R2\le 0.45$ (0.35)  
& 6\%         & ---  & &       \\
  Signal Extraction           & 5\%         & 13\%   & 8\%  & 8\%    \\
  Event generator mismodeling$^*$     &  10\%         & 10\%   & 10\% & 
10\%    \\ \\
  \multicolumn{3}{l}{\em Lambda Requirements$^*$}                         \\
  Lambda momentum: 0.8--1.2 GeV/c ({\bf 1.0 GeV/c}) & 6\% & 6\% & 6\% & 6\% \\
 Radial cut on $\Lambda$ vertex: 1.6--2.4 cm ({\bf 2.0 cm}) & 3\% & 3\% 
& 3\% & 3\%    \\
%  Vertex-finding                        & 2\% & 2\% & 2\% & 2\% \\
  \multicolumn{3}{l}{\em Additional Systematics}                        \\
  Tracking-Finding 
Uncertainty$^*$                     & 4\%  & 4\%  & 4\% & 4\%       \\
  D0 Cross Check                     & 5\%     & 3\%  & 6\%  & 14\%  \\ \hline
  Total Systematic Uncertainty     & 20 \% & 20\% & 15\% & 20\%
\\ \hline
  \multicolumn{5}{c}{\em RMS spread of tag values} 14\% \\ \hline
\end{tabular}
\end{table}

\end{document}